\documentstyle[twocolumn,epsf]{jpsj}
\def\runtitle
{Field Dependence of Electronic Specific Heat 
in Two-Band Superconductors}
\def\runauthor{Noriyuki {\sc Nakai}, 
Masanori {\sc Ichioka} and Kazushige {\sc Machida}}

\title
{Field Dependence of Electronic Specific Heat 
in Two-Band Superconductors}

\author
{ 
Noriyuki {\sc Nakai}\footnote
{E-mail:
nakai@mp.okayama-u.ac.jp},
Masanori {\sc Ichioka}
and Kazushige {\sc Machida}
}

\inst
{Department of Physics, Okayama University,
         Okayama 700-8530
}

\recdate{\today}

\abst{
The vortex structure is studied in light of MgB$_2$ theoretically
based on a two-band superconducting model by means of Bogoliubov-de Gennes
framework. The field dependence of the 
electronic specific heat coefficient $\gamma (H)$
is focused. The exponent $\alpha$ in $\gamma (H)\propto H^{\alpha}$ is shown 
to become smaller by adjusting the gap ratio of the two gaps on the major and minor bands.
The observed extremely small value $\alpha\sim 0.23$ 
could be reasonable in this two-band model with the gap ratio $\sim 0.3$.
}

\kword{%
MgB$_2$, two-band superconducting model, 
two gaps, vortex structure, electronic specific heat coefficient $\gamma (H)$, 
Bogoliubov-de Gennes equations
}

\begin{document}
\sloppy
\maketitle

Much attention has been focused on the recently discovered MgB$_2$
because of its relatively high superconducting transition temperature $T_{\rm c}\sim $ 39K
and simple crystalline structure\cite{nagamatsu}. General consensus obtained so far 
is that the electron-phonon interaction is mainly responsible for the pairing 
mechanism in this system because the large isotope effect is observed\cite{budko,hinks}.
There is, however, little consensus as to the 
microscopic description for the record high $T_{\rm c}$  due to the electron-phonon 
interaction.

Apart from the much debated pairing mechanism, it is rather urgent to determine 
the precise pairing function or gap function realized in MgB$_2$.
There are several important, but conflicting experimental data concerning
 the superconducting energy gap $\Delta$, ranging from the strong electron-phonon coupling  
$2\Delta/k_{\rm B}T_{\rm c}\sim5$ to extremely weak coupling value 
$2\Delta/k_{\rm B}T_{\rm c}\sim2$.
These come from the earlier experiments, such as position-dependent tunneling or Raman
experiments\cite{yamashita}. More recent experiments show unequivocally that these two
gap values come from a single sample and converge to definite values: the larger  $\Delta_L$
and the smaller $\Delta_S$ whose ratio $\Delta_S/\Delta_L$ falls around 0.3$\sim$0.4.
These experiments include photoemmision 
($\Delta_L$=5.6meV, $\Delta_S$=1.7meV, $\Delta_S/\Delta_L$=0.30)\cite{tsuda}, 
the $T$-dependent 
specific heat analysis ($\Delta_S/\Delta_L$=0.27)\cite{fisher,bouquet} and 
tunneling experiment ($\Delta_S/\Delta_L$=0.42)\cite{giubileo}.
Through these analyses, they are able to obtain systematic and 
smooth $T$ evolutions
of each gap value.
This implies that the two gap structure is an intrinsic property in MgB$_2$.

According to the band structure calculations\cite{kortus,yildirim}, there are two distinctive 
Fermi surface sheets; one is a two-dimensional cylindrical Fermi surface arising
from $\sigma$-orbitals due to $p_x$ and $ p_y$ electrons of B atoms and the other is a Fermi
surface coming from $\pi$-orbitals due to $p_z$ electrons of B atoms.
They are weakly hybridized with electron orbitals of Mg atoms.
Since the $\sigma$-orbital is strongly coupled to the in-plane B-atom vibration
with E$_{2g}$ symmetry simply because the hopping integral between the $\sigma$-orbitals
is modulated by this bond stretching motion. On the other hand, it is shown 
by the band calculation\cite{yildirim} that the $\pi$-orbital is weakly coupled 
with this phonon mode.
Thus it is quite conceivable that these two Fermi surfaces 
with different electronic characters have different energy-gap values
if this particular in-plane vibrational mode is responsible for the attractive 
interaction which induces superconductivity in MgB$_2$.
 
Here we are going  to analyze the field dependence of the $T$-linear
electronic specific-heat coefficient $\gamma(H)$ in the superconducting mixed
state by investigating the vortex lattice structure in two-band superconductors.
It is known that $\gamma(H)$ is a sensitive and useful quantity to reflect the gap
structure through the zero-energy excitation spectrum inside and outside the 
vortex core\cite{volovik,hasegawa1,hasegawa2,wang}. In particular, the exponent $\alpha$
in $\gamma(H)\propto H^{\alpha}$ at low fields reflects the nodal structure
of the superconducting gap
at the Fermi surface, playing a vital role to identify the gap function\cite{sonier}.
Several recent specific heat experiments on  MgB$_2$ show a very small 
exponent\cite{fisher,junod,yang} $\alpha\sim0.23$, implying that on increasing $H$ 
the zero-energy density of states (DOS) in the mixed state quickly recovers its normal
state value, 
compared with those in $d$-wave superconductors\cite{volovik,hasegawa1,hasegawa2,wang} 
with $\alpha\sim 0.5$ or clean limit $s$-wave case\cite{hasegawa1,hasegawa2} with 
$\alpha \sim 0.7$. Since there are no definitive reports which claim a line or point node
in the superconducting gap in  MgB$_2$, this small exponent remains mystery and requires
a proper explanation by  microscopic calculations. This is one of our purposes in this paper
based on the microscopic theory of Bogoliubov-de Gennes (BdG) framework.\cite{wang,takigawa1,takigawa2}

 We start with a  model pairing Hamiltonian for a two-band superconductor
 described by tight binding form:
\begin{eqnarray}
 {\hat H}&=&{\hat H}_0+{\hat H}_{pair} ,\\
 {\hat H}_0&=&\sum_{i,j,\sigma,\gamma}(-\tilde{t}_{i j \gamma}-\mu_{\gamma}\delta_{i, j})
 a^{\dagger}_{i \sigma \gamma}a_{j \sigma \gamma} ,\\
 {\hat H}_{pair}&=&\frac{1}{2}\sum_{i,\sigma,\gamma, \gamma'}g_{\gamma\gamma'}
 (a_{i \sigma \gamma}a_{i -\sigma \gamma})^{\dagger}
 a_{i \sigma \gamma'}a_{i -\sigma \gamma'}
\end{eqnarray}

\noindent
with the nearest neighbor (NN) hopping integral
  
\begin{eqnarray}
\tilde{t}_{i j \gamma}&=&
t_{\gamma}{\rm exp}[{\rm i}\frac{\pi}{\phi_0}
\int^{{\mib r}_j}_{{\mib r}_i}{\mib A}({\mib r})\cdot {\rm d}{\mib r}] ,
\end{eqnarray}

\noindent
where ${\mib A}({\mib r})$ is the vector potential and $\phi_0={hc/2e}$ 
 is the unit flux. The two-dimensional square lattice whose lattice constant is unity
 is assumed. The index $\gamma$ denotes the two bands $\gamma=L$ and $S$.
 Assuming the singlet pairing, we can derive the BdG equations
 for $\gamma=L$ and $S$ in a standard way:
 
\begin{eqnarray}
\sum_{i}
\left(
\begin{array}{@{\,}cc@{\,}}
K_{ji\gamma} & \delta_{i,j}\Delta_{i\gamma} \\
\delta_{i,j}\Delta_{i\gamma}^{\dagger} & -K_{ji\gamma}^{\ast}
\end{array}
\right)
\left(
\begin{array}{@{\,}c@{\,}}
u_{\gamma \epsilon}({\mib r}_i)\\
v_{\gamma \epsilon}({\mib r}_i)\nonumber 
\end{array}
\right)
&=&
E_{\gamma \epsilon}
\left(
\begin{array}{@{\,}c@{\,}}
u_{\gamma \epsilon}({\mib r}_j)\\
v_{\gamma \epsilon}({\mib r}_j)
\end{array}
\right)\\
\end{eqnarray}
\noindent
where 
\begin{eqnarray}
 K_{i j\gamma}=-\tilde{t}_{i j \gamma}-\mu_{\gamma}\delta_{i, j}.
\end{eqnarray}

\noindent
 The gap equation is given by 
\begin{eqnarray}
\Delta_{i \gamma}=\sum_{\gamma'}g_{\gamma\gamma'}d_{\gamma '}({\mib r}_i)
\end{eqnarray}

 \noindent
 with the order parameter 
\begin{eqnarray}
 d_{\gamma}({\mib r}_i)&=&\langle a_{i\downarrow\gamma}
 a_{i\uparrow\gamma}\rangle \nonumber \\
&=&-\sum_{\epsilon}v^{\ast}_{\gamma \epsilon}({\mib r}_i)
u_{\gamma \epsilon}({\mib r}_i)
\tanh {E_{\gamma \epsilon}\over 2T}.
\end{eqnarray}

\noindent 
 The local density of states (LDOS) at site $i$ for the $\gamma$ band is calculated 
 by 
\begin{eqnarray}
N_{\gamma}({\mib r}_i,E)=\sum_{\epsilon}\{|u_{\gamma \epsilon}({\mib r}_i)|^2
\delta(E-E_{\gamma \epsilon})&\nonumber\\
    +|v_{\gamma \epsilon}({\mib r}_i)|^2
\delta(E+E_{\gamma \epsilon})&\}.
\end{eqnarray}

 \noindent
 We assume an isotropic $s$-wave pairing for both bands $\gamma=L$ and $S$ characterized 
 by the order parameters (the energy gaps) $d_L(\Delta_L)$ and $d_S(\Delta_S)$.
 The attractive interactions are chosen as $g_{LL}\neq0$, $g_{LS}=g_{SL}\neq0$ and 
 $g_{SS}=0$, namely in eq. (5) the gap $\Delta_S$ on the $S$-band 
 is induced by the Cooper pair tunneling via $g_{LS}$.
 As for the normal state  band parameters we take $t_L=t_S=t(\equiv 1)$
 and $\mu_L=-1$ and  $\mu_S=+1$, thus the Fermi surface for $\gamma=L(S)$
 is close (open) around the $\Gamma$-point. 
The DOS for both bands is same at the Fermi level.
As two vortices are accommodated 
 in a unit cell of ${N_{\rm a} \times N_{\rm a}}$ atomic sites, the applied magnetic field is given by 
 $H_{N_{\rm a} {\times} N_{\rm a}}\equiv2\phi_0/N_{\rm a}^2$. 
By introducing the quasi-momentum of the magnetic Bloch state we obtain the wave function 
under the periodic boundary condition for a large number of unit cells
(See detailed numerical calculations in ref. 19).

 First, we study the case $g_{LL}=2.0$ and $g_{LS}=0.6$, which gives 
$\Delta_L=0.322$ and $\Delta_S=0.086$ at zero field. 
It is designed to adjust the gap ratio for MgB$_2$($\Delta_S/\Delta_L=0.27$). 
If we consider a single band superconductor with the small gap $\Delta=0.086$, 
the superconductivity vanishes at the following magnetic field discussed below, 
since $H > H_{{\rm c}2}$. But, in this two band superconductor, 
the small gap superconductivity survives in the $S$-band 
because of the cooper pair transfer $g_{LS}$.
 
\begin{figure}[t]
\begin{center}
\epsfbox{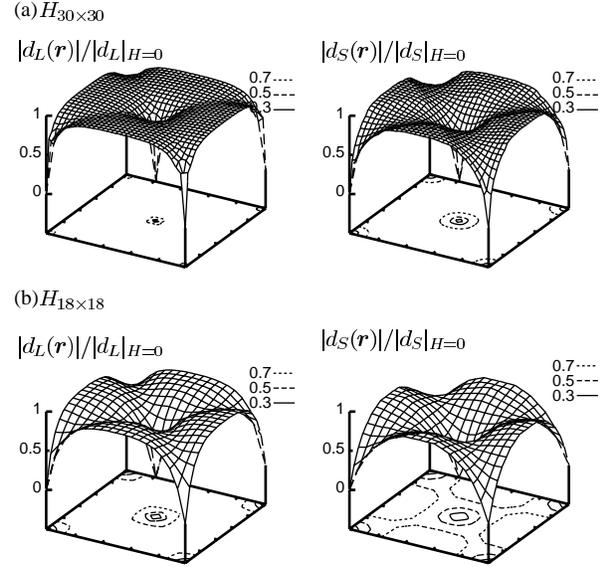}
\end{center}
\caption{
Spatial profiles of the order parameter $|d_L({\mib r})|$ for the $L$-band 
and $|d_S({\mib r})|$ for the $S$-band 
at lower field $H_{30 \times 30}$(a) and at higher field $H_{18 \times 18}$(b). 
They are normalized by the zero field values 
$|d_L|_{H=0}(=0.145)$ and $|d_S|_{H=0}(=0.055)$, respectively.
}
\label{odp}
\end{figure}
\begin{figure}[tb]
\begin{center}
\epsfbox{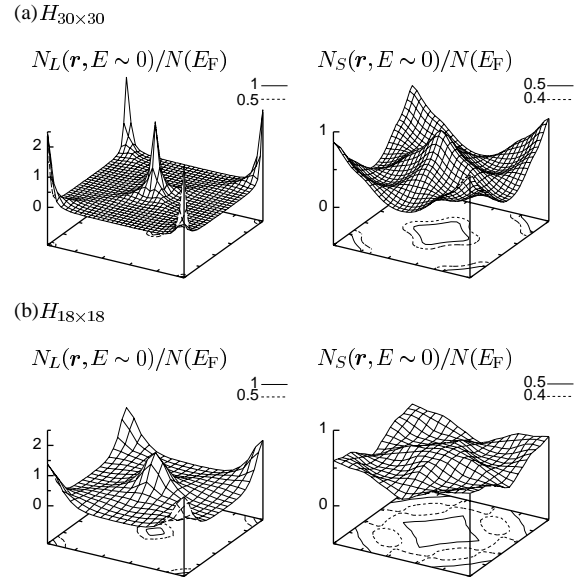}
\end{center}
\caption{
Zero energy local density of states ${N_L({\mib r}, E\sim0)}$ 
for the $L$-band and ${N_S({\mib r}, E\sim0)}$ for the $S$-band 
at lower field $H_{30 \times 30}$(a) and at higher field $H_{18 \times 18}$(b). 
They are normalized by $N(E_{\rm F})$, the normal state DOS at the Fermi level.
}
\label{ldp}
\end{figure}
 The spatial profiles of the order parameters $d_L({\mib r})$ and $d_S({\mib r})$
are shown in Fig. 1 where the unit cell of the square vortex lattice is displayed. 
Vortices are accommodated at the center and four corners. 
It is seen that the vortex core radius 
for the $L$-band ($S$-band) is small (large) and the depression of $|d({\mib r})|$
is apparent along the NN direction, in particular, for the $S$-band. 
By increasing $H$, $|d({\mib r})|$ is further suppressed as is seen in Fig. 1 where 
the core radius is widen. 
The suppression by $H$ is eminent in the $S$-band. 

The corresponding spatial profiles of the LDOS are shown in Fig. 2, where 
$N_L({\mib r}, E\sim0)$ and  $N_S({\mib r}, E\sim0)$ have a peak 
at the vortex center 
and the ridges connecting the vortex cores are clearly seen.
While the high density of states is concentrated 
at the vortex core in ${N_L({\mib r},E\sim0)}$, it rather
spreads out in ${N_S({\mib r},E\sim0)}$. 
This is because the vortex bound states are highly confined in
the $L$-band vortex corresponding to the narrow core radius while 
in the $S$-band vortex the core states are loosely bounded. 
The spatial profiles for $N_L({\mib r},E\sim0)$ and $N_S({\mib r},E\sim0)$ are 
resemble to those of the low-field case and the high-field case 
in the single band superconductor (see Fig. 1 and Fig. 2 in ref. 13). 
In $N_S({\mib r},E\sim0)$, the low energy states extending from vortex cores 
overlap with each other, and the LDOS is suppressed along the line 
connecting the NN or next NN vortices. With increasing $H$, 
the effect by the overlap becomes eminent, and the LDOS is reduced to 
the flat profile $N_S({\mib r},E\sim0)/N(E_{\rm F})\sim0.5$ in the $S$-band
($N(E_{\rm F})$ is the total DOS in the normal state at the Fermi level). 

The spatial average of 
 ${N_L({\mib r},E\sim0)}$ and  ${N_S({\mib r},E\sim0)}$ gives rise to the total DOS under a given field, leading
 to ${\gamma(H)}$ which is defined by
\begin{eqnarray}
 \gamma(H)=\gamma_L(H)+\gamma_S(H)
\end{eqnarray}
 with 
\begin{eqnarray}
\gamma_{L,S}(H)=\langle N_{L,S}({\mib r}, E\sim0)\rangle _{{\mib r}:{\rm unit\, cell}}.
\end{eqnarray}
\begin{figure}[t]
\begin{center}
\epsfbox{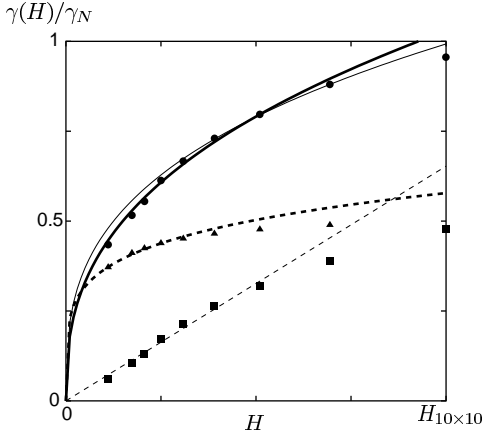}
\end{center}
\caption{
Field dependence of ${\gamma(H)}$ for ${\Delta_S/\Delta_L=0.27}$. 
Points of $\gamma(H)$(circles), $\gamma_{S}(H)$(triangles) and $\gamma_L(H)$(squares) 
are numerical data. 
The thick line is fitting for lower field data of $\gamma(H)$.
The thin line is fitting by ${\gamma(H)\sim\gamma_{N}(H/H_{{\rm c}2})^{\alpha}}$. 
In the low field the thick(thin) dotted line is fitting for $\gamma_S(H)$($\gamma_L(H)$). 
}
\label{gamma1}
\end{figure}%
 We have done extensive computations for various $H_{N_{\rm a} \times N_{\rm a}}$cases. 
In Fig. 3, 
it is seen that $\gamma(H)$ is described by a power law:
 $\gamma(H)\propto H^{\alpha}$ with small $\alpha$.
If only the low field points are fitted, we obtain ${\alpha=0.38}$(thick line). 
The fitting by ${\gamma(H)\sim\gamma_{N}(H/H_{{\rm c}2})^\alpha}$ 
under the condition that $\gamma(H)$ is reduced to the normal state value $\gamma_N$
gives $\alpha =0.33$(thin line). The small exponents $\alpha$, 
 or the sharp rise of $\gamma(H)$ in small fields, can be attributed to
the $S$-band contribution $\gamma_S(H)\propto H^{0.20}$, 
while $\gamma_L(H)\propto H^{1.00}$ in the $L$-band.
That is, the small $\alpha$ is due to the overlap of the low energy states 
outside of vortex cores at the $S$-band.
 Physically it is because the energy gap for the $S$-band is suppressed by a weak field, 
while the total superconductivity is maintained by the larger energy 
 gap up to $H_{{\rm c}2}$. This intuitively appealing picture is actually confirmed by the 
 present microscopic calculation. This is, however, different from the two independent gaps 
 with different transition temperatures and different $H_{{\rm c}2}$. In such a case, we would have 
 double transitions and $\gamma(H)$ would be a simple addition of two independent curves, 
which has a kink structure at the lower $H_{{\rm c}2}$. This is not the case for MgB$_2$.
 
\begin{figure}[tb]
\begin{center}
\epsfbox{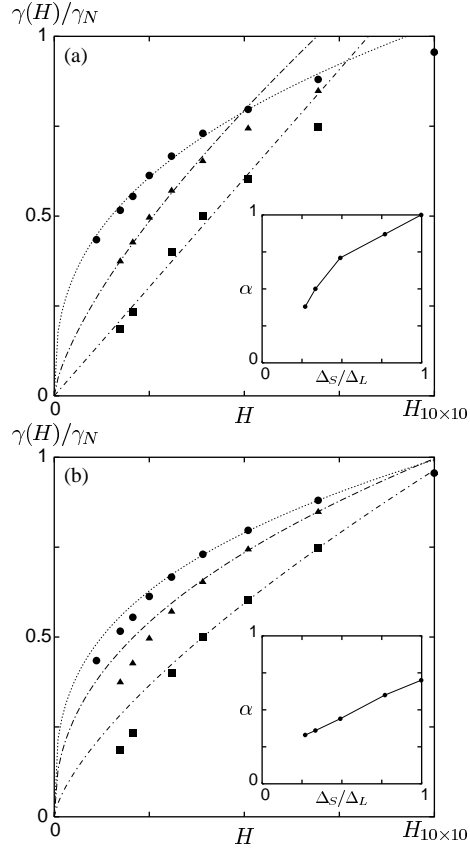}
\end{center}
\caption{
Field dependence of $\gamma(H)$ for $\Delta_S/\Delta_L=0.27$(circles), 
0.49(triangles) and 0.997(squares). (a)Fitting lines for the low field data. 
(b)Fitting lines by $\gamma(H)\sim \gamma_N(H/H_{\rm c})^{\alpha}$. 
Points of numerical data are the same in both figures. 
In the insets, $\Delta_{S}/\Delta_{L}$-dependence of $\alpha$ 
is shown for each fitting case.
}
\label{gamma2}
\end{figure}%
To study the dependence of $\alpha$ on the gap ratio $\Delta_S/\Delta_L$, 
we perform the calculation for various pairing parameter sets. 
In Fig. 4, we show the $\Delta_S/\Delta_L$-dependence of $\gamma(H)$ behavior. 
There, we show the results using the two kinds of fitting; 
the fitting for low field data in Fig. 4(a), 
and overall fitting by ${\gamma(H)\sim\gamma_{N}(H/H_{{\rm c}2})^{\alpha}}$ in Fig. 4(b), 
while the numerical data are the same in both figures. In the insets, 
we show the $\Delta_S/\Delta_L$-dependence of $\alpha$ in each fitting case. 
In the limiting case $\Delta_S/\Delta_L\to 1$, $\alpha$ is reduced to the exponent in the 
single band case. In Fig. 4(b), it gives $\alpha\sim0.7$ 
in accord with the previous quasi-classical calculation.\cite{hasegawa1}
In both cases, $\alpha$ monotonically decreases with decreasing $\Delta_S/\Delta_L$. 
It should be emphasized that we may identify the gap ratio ${\Delta_S/\Delta_L}$ uniquely
 by measuring the electronic specific heat under varying external field,
 providing a rather unique spectroscopic method for determining the gap ratio. 
In Fig. 4, for the gap ratio ${\Delta_S/\Delta_L}\sim0.3$ observed 
by the several groups with different methods, we obtain small exponent $\alpha$, 
as in the specific heat data on MgB$_2$.

There are several factors which might alter our conclusion on the relation $\alpha$
 vs. ${\Delta_S/\Delta_L}$.
 
 (1) We assume that the DOS in the normal state for each Fermi
 surface sheet is equal. According to the band calculation by Belashchenko
 {\it et al.}\cite{bela} the ratio of the two DOS is 0.55 ($\pi$-band) : 0.45 ($\sigma$-band).
 This small difference causes potentially to alter our conclusion, but not in an essential
 way.
 
 (2) It is assumed that in the minor $S$-band there is no direct attractive 
 interaction $g_{SS}=0$. The gap in the $S$-band is exclusively induced by
 the Cooper pair tunneling process via $g_{LS}\neq0$.
According to Kortus {\it et al}\cite{kortus}, the electron-phonon coupling in the 
$S$-band due to the bond stretching mode is smaller than 
that in the $L$-band, but yet non-vanishing. Thus  $g_{SS}$ might not completely
vanish in MgB$_2$. We might regard it vanish as a first approximation
because our conclusion relies exclusively on the gap ratio ${\Delta_S/\Delta_L}$.
The effect of $g_{SS}\neq0$ will be studied in the future for more details.

(3) We comment on the small discrepancy of the exponent $\alpha$ 
between our calculation based on BdG theory and that of the experiment. 
Since our parameters belong to a rather quantum limit case
(the coherent length $\xi$ is an order of the atomic lattice constant), the 
quasi-classical calculation\cite{hasegawa1,hasegawa2} is more appropriate for MgB$_2$.
We believe, however, that the overall relation   $\alpha$ vs.  ${\Delta_S/\Delta_L}$ is 
not greatly altered in that calculation. We will study this case in a future publication.

Haas and Maki\cite{haas} analyze an anisotropic $s$-wave pairing state in connection 
with MgB$_2$. Their single band model is designed to describe the anisotropic 
superconducting properties such as the upper critical field or the penetration depth. 
According to the recent penetration depth measurement\cite{manzano} for single crystals of  
MgB$_2$, the anisotropy of the penetration depths for $H\parallel c$ and $H\perp c$ 
in the hexagonal crystal is almost absent, which is at odd with the prediction by Haas and
Maki. Since their single band model is similar to our two band model in the
sense that the gap anisotropy is implemented in the reciprocal space in 
Haas and Maki or implemented in the energy space in ours.
In order to fully describe the three dimensional superconducting nature in MgB$_2$
our model should consider the anisotropic $s$-wave pairing state for both major and minor bands,
which may better explain the above penetration depth experiment.

  We speculate that the present multi-gap model may have potentially wide applicability.
  It is quite usual that a superconductor has a multiple gap because the underlying
  Fermi surface consists of multiple sheets, on each of which the gap value
  could be different. It is true even for elemental metals. MgB$_2$ may belong
  to an extreme case. To reveal this feature, the measurement of $\gamma(H)$
  is demonstrated to be a useful tool. 
The analysis for the $p$-wave pairing case\cite{takigawa3}, 
focusing on Sr$_2$RuO$_4$, is reported in ref. 24.

In conclusion, we have evaluated the exponent  $\alpha$ in the $T$ linear 
specific heat coefficient $\gamma(H)\propto H^{\alpha}$ for a simple two-band
superconductor and succeeded in reproducing the extremely small $\alpha\cong0.3$, 
as in observed in MgB$_2$, by taking the two gap ratio $\Delta_S/\Delta_L\cong0.3$, each
coming from the different Fermi
sheets. Thus we conclude that the gap functions are distinctly different for the 
different Fermi sheets, the major is the $\sigma$-band ($p_x$ and $p_y$ characters of B atoms)
 while the minor is the $\pi$-band ($p_z$ characters of B atoms), yet each gap
 being isotropic on its own Fermi sheet. Thus we do not need exotic anisotropic
 gap function for describing superconductivity here.
 This two-band feature is intrinsic in MgB$_2$.

\end{document}